\documentclass[aps,prl,twocolumn,superscriptaddress,showpacs,showkeys]{revtex4}

\usepackage{graphicx,color}
\newcommand{\fatq}{\mathbf{q}}
\newcommand{\fatQ}{\mathbf{Q}}

\newcommand{\pgnfigure}[2]{\begin{figure}[h]\includegraphics[width=7.4cm]
{#1}\caption{\label{#1}#2}\end{figure}}

\definecolor{mycolor}{rgb}{0.8, 0.2, 0.2}

\begin{document}

\title{Ultra-small-moment incommensurate spin density wave order masking a ferromagnetic quantum critical point in NbFe$_2$}

\author{P. G. Niklowitz}
\email[e-mail:]{philipp.niklowitz@rhul.ac.uk}
\affiliation{Department of Physics, Royal Holloway, University of
London, Egham TW20 0EX, U.K.}

\author{M. Hirschberger}
\affiliation{Department of Physics, Princeton University, Princeton, NJ08544, U.S.A.}

\author{M. Lucas}
\affiliation{Department of Physics, Royal Holloway, University of
London, Egham TW20 0EX, U.K.}

\author{P. Cermak}
\affiliation{Charles University, Faculty of Mathematics and Physics, Department of Condensed Matter Physics, Ke Karlovu 5, 121 16, Praha, Czech Republic}

\author{A. Schneidewind}
\affiliation{J\"ulich Centre for Neutron Science (JCNS) at Heinz Maier-Leibnitz Zentrum (MLZ), Forschungszentrum J\"ulich GmbH, Lichtenbergstr. 1, 85748 Garching, Germany}

\author{E. Faulhaber}
\affiliation{Heinz Maier-Leibnitz Zentrum (MLZ), Technische Universit\"at M\"unchen, Lichtenbergstr. 1, 85748
Garching, Germany}

\author{J.-M. Mignot}
\affiliation{Laboratoire L\'eon Brillouin (CEA-CNRS), CEA Saclay, F-91911
Gif-sur-Yvette, France}

\author{W. J. Duncan}
\affiliation{Department of Physics, Royal Holloway, University of
London, Egham TW20 0EX, U.K.}

\author{A. Neubauer}
\affiliation{Physik Department E21, Technische Universit\"at
M\"unchen, 85748 Garching, Germany}

\author{C. Pfleiderer}
\affiliation{Physik Department E21, Technische Universit\"at
M\"unchen, 85748 Garching, Germany}

\author{F. M. Grosche}
\affiliation{Cavendish Laboratory, University of Cambridge, Cambridge CB3 0HE, U.K.}

\begin{abstract}
In the metallic magnet Nb$_{1-y}$Fe$_{2+y}$, the low temperature threshold of ferromagnetism can be investigated by varying the Fe excess $y$ within a narrow homogeneity range. We use elastic neutron scattering to track the evolution of magnetic order from Fe-rich, ferromagnetic Nb$_{0.981}$Fe$_{2.019}$ to approximately stoichiometric NbFe$_2$, in which we can, for the first time, characterise a long-wavelength spin density wave state burying a ferromagnetic quantum critical point. The associated ordering wavevector $\fatq_{\rm SDW}=$(0,0,$l_{\rm SDW}$) is found to depend significantly on $y$ and $T$, staying finite but decreasing as the ferromagnetic state is approached. The phase diagram follows a two order-parameter Landau theory, for which all the coefficients can now be determined. Our findings suggest that the emergence of SDW order cannot be attributed to band structure effects alone. They indicate a common microscopic origin of both types of magnetic order and provide strong constraints on related theoretical scenarios based on, e.g., quantum order by disorder.
\end{abstract}

\pacs{[75.25.-j, 75.40.-s, 75.40.Cx, 75.50.Bb]}
\keywords{ferromagnetism, quantum phase transitions, quantum criticality, new phases, emerging modulated order, intermetallic systems}

\maketitle

%%%%%%%%%%%%%%%%%%%%%%%%%%%%%%%%%%%%%%%%%%%%%%%%%%%%%%%%%%%%%%%%%%%%%%%%%

The exploration of ferromagnetic quantum phase transitions in metals has motivated numerous theoretical and experimental studies \cite{bra16a}, which have led to the discovery of non-Fermi liquid states \cite{uhl04a,nik04d} and of unconventional superconductivity (e.g. \cite{pfl02a,sax00a,ali10a}). The underlying question, however, whether a ferromagnetic quantum critical point (QCP) can exist in clean band magnets, remains controversial. Fundamental considerations \cite{pfl09a,voj99a,chu04a} suggest that the ferromagnetic QCP is avoided in clean systems by one of two scenarios: either the transition into the ferromagnetic state becomes discontinuous (first order), or the nature of the low temperature ordered state changes altogether, for instance into nematic or long-wavelength spin density wave (SDW) order \cite{voj99a,chu04a}. Whereas there are many examples for the first scenario, including ZrZn$_2$ \cite{uhl04a}, Ni$_3$Al \cite{nik04d} and UGe$_2$ \cite{pfl02a}, the transition into a modulated state on the border of band ferromagnetism has proven to be more challenging to investigate. Recent reports show that this scenario may apply more widely beyond the comparatively simple band ferromagnets for which it was first discussed: (i) the masking of the field-tuned quantum-critical end point of the continuous metamagnetic transition of Sr$_3$Ru$_2$O$_7$ by two SDW phases \cite{les15a}, (ii) the evolution of FM into long-wavelength SDW fluctuations in the heavy-fermion system YbRh$_2$Si$_2$ \cite{sto12a}, which displays a high Wilson ratio \cite{geg05a} and 
becomes FM under Co-doping \cite{lau13a}, (iii) the emergence at finite temperature of SDW order in the ferromagnetic local moment system PrPtAl \cite{abd15a}, and the appearance of modulated magnetic order at the border of pressure-tuned FM systems CeRuPO \cite{len15a}, MnP \cite{che15a}, or LaCrGe$_3$ \cite{kal17a}.

The band magnet NbFe$_2$ is a particularly promising candidate for investigating the SDW scenario in a clean itinerant system, because it is located near the border of ferromagnetism at ambient pressure \cite{shi87a}, enabling multi-probe studies and, in particular, neutron scattering. Ferromagnetic order can be induced at low temperature by growing Fe-rich Nb$_{1-y}$Fe$_{2+y}$ with $y$ as small as 1\% (Figure \ref{figure_1} \cite{mor09a}). Compton scattering results on the Fe-rich side of the phase diagram have been analyzed by assuming ferrimagnetism as the ground state \cite{hay12a}, but more direct probes of the local fields by M{\"o}{\ss}bauer spectroscopy point to ferromagnetism as the ground state \cite{rau15a}. The precise low temperature state for $y=0$ has remained unidentified since early NMR studies first suggested that stoichiometric NbFe$_2$ may display low-moment SDW order \cite{yam88a}. Repeated attempts to detect the SDW order in neutron scattering were unsuccessful, but recent results from ESR, $\mu$SR, and M\"o{\ss}bauer spectroscopy also point strongly towards SDW order \cite{rau15a}. Non-Fermi liquid forms of resistivity and low temperature heat capacity have been observed in slightly Nb-rich NbFe$_2$. \cite{bra08a}

\pgnfigure{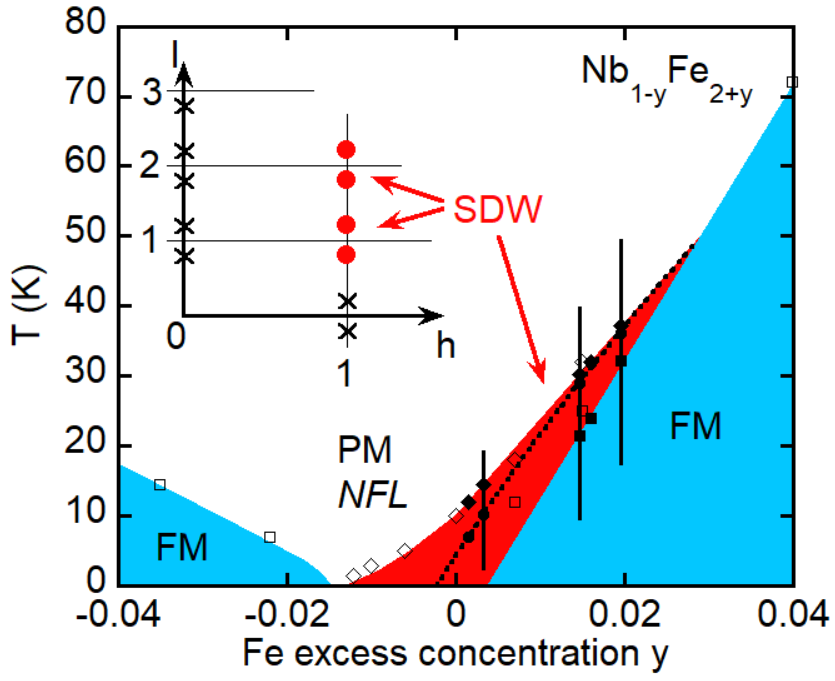}{Phase diagram of Nb$_{1-y}$Fe$_{2+y}$ with results for bulk $T_{\rm C}$ (squares) and $T_{\rm N}$ (diamonds) from single-crystal neutron diffraction (filled symbols) embedded into previous results from polycrystals (empty symbols)\cite{mor09a}. Vertical solid lines indicate the $T$ range of neutron diffraction measurements. Of the two ferromagnetic (FM) phases, the one on the more Fe-rich side is separated from the paramagnetic (PM) state by a spin-density wave (SDW) at low temperatures, where non-Fermi liquid (NFL) behaviour is found as well. $T_{\rm 0}$, the FM phase boundary buried by the SDW phase (dashed line) is an extrapolation of $T_{\rm 0}$ values (circles) measured or calculated for the single crystals \cite{nik19a_sup}. The inset shows the relevant reciprocal-space region, which was accessible during the neutron diffraction experiments. Circles show the presence and crosses the absence of SDW peaks. The SDW peak pattern is consistent with moments pointing along the $c$-axis.}

\begin{figure*}\includegraphics[width=\textwidth]
{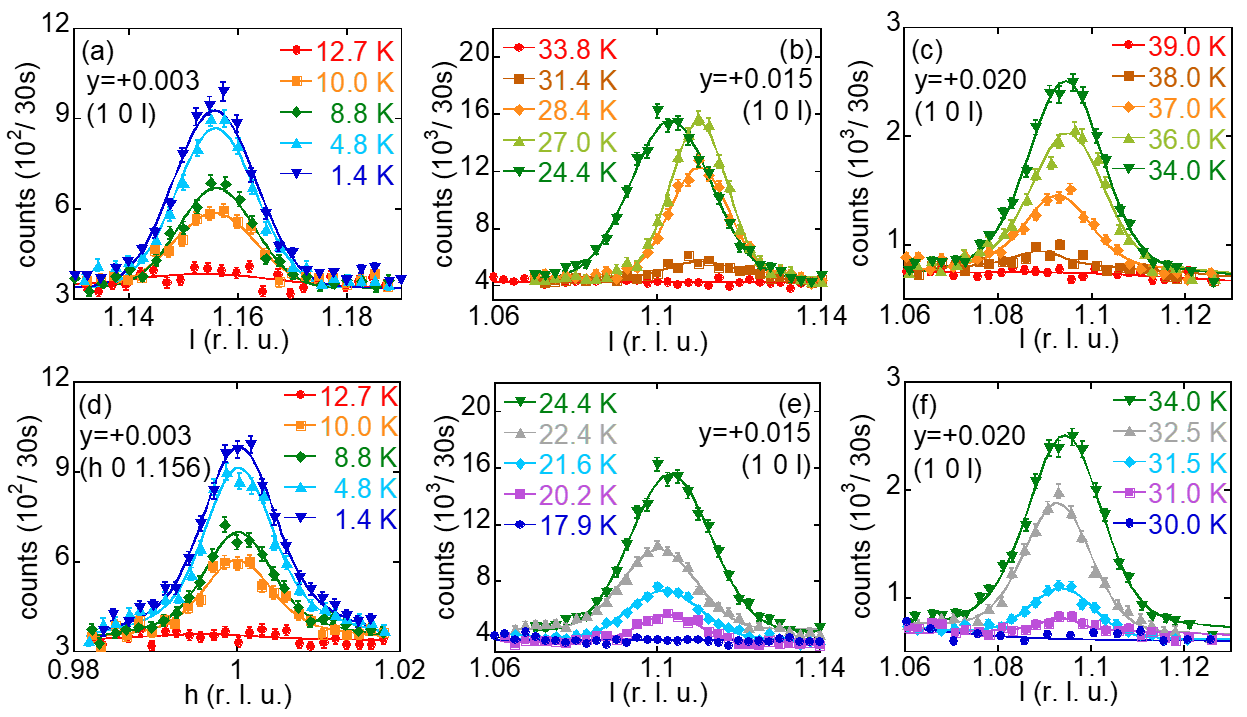}\caption{\label{figure_2} Neutron diffraction at $\fatQ=(1,0,1+l_{\rm SDW})$ of Nb$_{1+y}$Fe$_{2-y}$ samples A ($y=+0.003$), B ($y=+0.015$), and C ($y=+0.020$). The figures contain selected $h$ and $l$ scans accross the SDW Bragg peak demonstrating its $T$ dependence in each sample. Gaussian fits of the data are shown as solid lines. $l_{\rm SDW}$ has significant $y$ and $T$ dependence and is located in the range $0.083(1)\leq l_{\rm SDW}\leq 0.164(3)$ in the explored samples. The scans selected for this figure have been measured in sequences going down in $T$. The whole collected data set (not all data shown) contains $h$ and $l$ scans from sequences going up and down in $T$ for all samples.}\end{figure*}

\pgnfigure{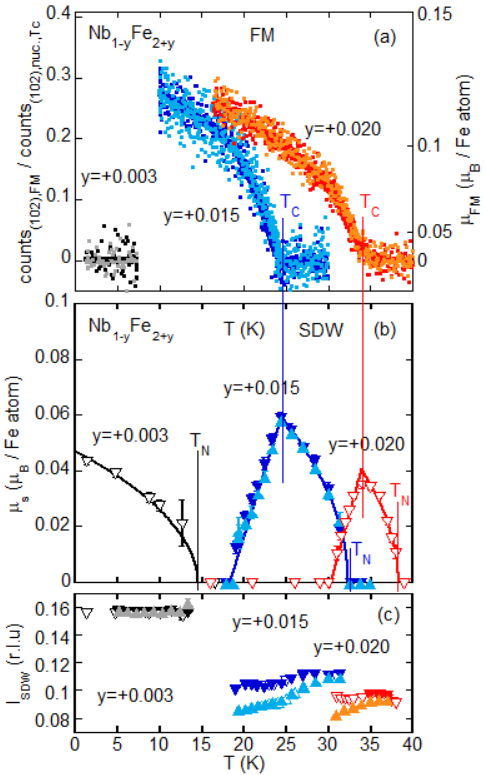}{$T$ dependences of normalised FM intensities and FM ordered moments $\mu_{\rm FM}$, SDW ordered moments $\mu_{\rm s}$, and of SDW ordering wavevector values $\fatq_{\rm SDW}$ of Nb$_{1-y}$Fe$_{2+y}$ obtained from measurement sequences going down (downward triangles) and up (upward triangles) in $T$ for Samples A ($y=+0.003$, black, grey), B ($y=+0.015$, dark blue, light blue), and C ($y=+0.020$, red, orange). Each sample has been investigated in two separate experiments (empty and filled symbols, respectively) demonstrating the reproducibility of the results. Lines are guides to the eye. (a) FM $(1 0 2)$ intensities have been normalised by the nuclear $(1 0 2)$ intensities at $T_{\rm C}$. The nuclear $(1 0 2)$ intensities have been subtracted. Conversion to FM ordered moments $\mu_{\rm FM}$ has been obtained as described in the Supplement \cite{nik19a_sup}. (b) SDW ordered moments $\mu_{\rm s}$ have been obtained from integrated intensities at $\fatQ=(1,0,1+l_{\rm SDW})$ and comparison with nuclear and FM $(1 0 2)$ intensities \cite{nik19a_sup}. Lines to the right of the maxima are fits described in \cite{nik19a_sup}. Lines to the left of the maxima are guides to the eye. (c) For all studied samples $\fatq_{\rm SDW}$ has the form (0,0,$l_{\rm SDW}$) in the whole $T$ range. A thermal hysteresis is observed in $l_{\rm SDW}$ of Samples B and C, which contain a SDW-FM phase transition.}

Here, we present the outcomes of a neutron diffraction study, which for the first time demonstrates unambiguously the existence of a long-wavelength modulated magnetic state (SDW) forming on the border of ferromagnetic order at low temperature in stoichiometric single crystals of NbFe$_2$. We track its evolution with temperature and composition and probe the underlying ferromagnetic order. We find that the SDW state indeed displays a very small ordered moment $\mu_{\rm s}<0.1$\,$\mu_{\rm B}/(\mbox{Fe atom})$, which explains why previous neutron scattering experiments failed to detect it. Our data confirms the second order nature of the PM-SDW phase transition and a temperature hysteresis in the SDW ordering wavevector suggests that the SDW-FM phase transition is first order. The observed characteristics of the SDW order including the evolution of its ordering wavevector, which we find decreases on approaching the FM transition, as well as our theoretical analysis of the resulting phase diagram suggest that the occurrence of long wavelength SDW order on the border of FM in NbFe$_2$ is not coincidental, but rather emerges from the proximity to a FM quantum critical point, which is buried within the SDW dome.

%%%%%%%%%%%%%%%%%%%%%%%%%%%%%%%%%%%%%%%%%%%%%%%%%%%%%%%%%%%%%%%%%%%%%%%%%

{\em Experimental.}---

Large single crystals of C14 Laves phase NbFe$_2$ (lattice constants $a=4.84$\,\AA\ and $c=7.89$\,\AA) with compositions chosen across the iron-rich side of the homogeneity range have been grown in a UHV-compatible optical floating zone furnace from polycrystals prepared by induction melting \cite{neu11a}. The single crystals have been characterised extensively by resistivity, susceptibility, and magnetisation measurements, as well as by x-ray diffraction and neutron depolarization \cite{pfl10a,neu11b,dun11a}, the latter showing homogeneity in structure and chemical composition. In this study, three samples have been measured: (i) sample A, which  is almost stoichiometric ($y = +0.003$); (ii) sample B, which is slightly Fe-rich ($y = +0.015$); and (iii) sample C, which is more Fe-rich still ($y = +0.020$) \cite{nik19a_sup}.

For the neutron scattering experiments the samples were mounted on Al holders and oriented with ($h0l$) as the horizontal scattering plane. In order to enhance the signal to background ratio neutron diffraction was carried out at two cold triple-axis spectrometers in elastic mode: Panda at the Heinz Maier-Leibnitz Zentrum (MLZ) \cite{hei15a} and 4F2 at the Laboratoire L{\'e}on Brillouin (LLB). Panda was run with neutron wavevectors $k_{\rm i}=k_{\rm f}=1.57$\,\AA$^{-1}$ and 4F2 was used with $k_{\rm i}=k_{\rm f}=1.30$\,\AA$^{-1}$ \cite{nik19a_sup}.  

%%%%%%%%%%%%%%%%%%%%%%%%%%%%%%%%%%%%%%%%%%%%%%%%%%%%%%%%%%%%%%%%%%%%%%%%%

{\em Results.}---  

The principal discovery of SDW Bragg reflections in the Nb$_{1-y}$Fe$_{2+y}$ system is presented in Fig.\,\ref{figure_2}, which focuses on the reflection $(1 0 1)^+$ at $\fatQ=(1,0,1)+\fatq_{\rm SDW}$ with $\fatq_{\rm SDW}=(0,0,l_{\rm SDW})$. SDW Bragg reflections corresponding to an ordering wavevector $\fatq_{\rm SDW}$ have been confirmed for all samples of this study. $l_{\rm SDW}$ shows a significant $y$ and $T$ dependence, which will be discussed further below. We have also established the SDW's long-range character \cite{nik19a_sup}. In addition to the data discussed above of the $(1 0 1)^+$ reflection, the vicinity of nuclear Bragg reflections has been scanned for further SDW satellite peaks. SDW satellite peaks are present at $\fatQ=(1,0,1)\pm\fatq_{\rm SDW}$ and $\fatQ=(1,0,2)\pm\fatq_{\rm SDW}$ but they are absent at $\fatQ=(1,0,0)\pm\fatq_{\rm SDW}$ and at $\fatQ=(0,0,l)\pm\fatq_{\rm SDW}$ with $l = 1,2,3$ (see inset of Fig.\,\ref{figure_1}). This distribution of allowed and forbidden satellite Bragg peaks is consistent with moment orientation along the $c$-axis, suggesting a lineary polarized SDW state rather than spiral order. Based on AC susceptibility and torque magnetometry data, the $c$-direction has been determined to be the magnetic easy-axis independent of the chemical composition \cite{kur97a,hir11a}.

Magnetic neutron scattering from the FM order has been observed at the position of the weak nuclear Bragg point $(1 0 2)$. Fig.~\ref{figure_3}a shows the $T$ dependence of the intensities of the FM $(1 0 2)$ Bragg reflections normalised by the intensities of the nuclear $(1 0 2)$ reflections at $T_{\rm C}$ for all three compositions. Conversion to FM ordered moments $\mu_{\rm FM}$ has been obtained as described in \cite{nik19a_sup}. In the Fe rich Samples B and C, FM order is observed at low temperatures. The measurements of the FM $(1 0 2)$ signals (Fig.~\ref{figure_3}a) and peaks of the SDW signals (Fig.~\ref{figure_3}b, see below) reveal onset temperatures at $T=24.5$\,K in sample B and $T=34$\,K in sample C.

The $T$ dependences of the SDW ordered moments $\mu_{\rm s}$ for all samples are shown in Fig.~\ref{figure_3}b. $\mu_{\rm s}$ has been obtained from integrated intensities at $\fatQ=(1,0,1+l_{\rm SDW})$ and comparison with nuclear and FM $(1 0 2)$ intensities as described in \cite{nik19a_sup}. In the almost stoichiometric Sample A, SDW order emerges below $T_{\rm N}=14.5$\,K and is present down to the lowest measured temperature of $1.4$\,K. In the slightly Fe-rich Sample B SDW order appears below $T=32.3$\,K and is fully suppressed below $T=18.5$\,K. Finally, in the most Fe-rich Sample C SDW order appears below $T=38.3$\,K and is fully suppressed below $T=30.5$\,K. The small size of the SDW moments ($\mu_{\rm s}<0.1$\,$\mu_{\rm B}/(\mbox{Fe atom})$) explains the difficulties in observing the SDW order in previous neutron diffraction experiments.

In all three samples, the SDW intensity rises continuously below the onset temperature $T_{\rm N}$, suggesting a second-order PM-SDW transition. The peak SDW intensity coincides with the FM onset temperature $T_{\rm C}$ in samples B and C, and there is a $T$-range below $T_{\rm C}$, in which SDW and FM order appear to coexist. This overlap can be attributed to a distribution of transition temperatures within the sample, giving bulk $T_{\rm N}=30.1$\,K, $T_{\rm C}=21.5$\,K, (Sample B) and $T_{\rm N}=37.1$\,K, $T_{\rm C}=32.2$\,K (Sample C) in good agreement with bulk magnetic response \cite{nik19a_sup}.

The SDW ordering wavevector $\fatq_{\rm SDW}=(0,0,l_{\rm SDW})$ is found to depend significantly on composition and temperature (Fig.~\ref{figure_3}c): (i) the $c$-axis pitch number shifts from $l_{\rm SDW}(T_{\rm N})=0.157(1)$ in Sample A to $l_{\rm SDW}(T_{\rm N})=0.095(1)$ in Sample C (corresponding to an incommensurate modulation along the $c$-axis with a pitch in the range $\lambda_{\rm SDW}\approx 50-100$\,\AA); (ii) in Samples B and C $l_{\rm SDW}$ shows a significant $T$ dependence, decreasing by about 20\% with decreasing $T$; (iii) $l_{\rm SDW}(T)$ stays finite at the SDW-FM transition, so there is a discontinuous change in the magnitude of the magnetic ordering wavevectors there from finite ($q_{\rm SDW}$) to zero ($q_{\rm FM}$); (iv) in Samples B and C $l_{\rm SDW}(T)$ reproducibly displays significant thermal hysteresis with lower $l_{\rm SDW}$ values when warming into the SDW phase from the FM state. (i) and (ii) means that $l_{\rm SDW}$ is being reduced on approaching FM and (iii) and (iv) point to the first-order nature of the SDW-FM transition.

%%%%%%%%%%%%%%%%%%%%%%%%%%%%%%%%%%%%%%%%%%%%%%%%%%%%%%%%%%%%%%%%%%%%%%%%%

{\em Discussion.}---

The delicate SDW state of the Nb$_{1-y}$Fe$_{2+y}$ system occurs in a narrow composition and temperature range attached to the threshold of FM order (Fig. \ref{figure_1}). Because the SDW-FM transition is first order, the emergence of the SDW state masks the FM QCP and buries it inside a SDW dome. Our neutron diffraction results have revealed the following main characteristics of the SDW state: (i) an incommensurate ordering wavevector $\fatq_{\rm SDW}$ with its striking dependence on $y$ and $T$, (ii) the long range nature in contrast to what would be expected for a spin glass \cite{bin86a}, and (iii) a small linearly polarised ordered moment. The small ordered moments, which contrast with the large fluctuating moments $\mu_{\rm eff}\approx1$\,$\mu_{\rm B}$ derived from the temperature dependence of the magnetic susceptibility, and the linearly polarized rather than helical order suggest that the SDW state should be understood within a band picture, not a local moment picture. However, the $y$ and $T$ dependent values of $\fatq_{\rm SDW}$ cannot be explained based on the electronic band structure alone.

Within density functional theory (DFT), the implications of the electronic band structure for magnetic order have been examined in detail. Direct total energy calculations for different ordering patterns \cite{sub10a,tom10a,nea11a} suggest that energy differences between a number of magnetic ground states are very small. Moreover, the wavevector dependence of the bare band structure derived susceptibility $\chi_{\fatq}^{(0)}$ or Lindhard function \cite{sub10a,tom10a} is not consistent with the long-wavelength SDW order reported here as it does not feature a significant enhancement near the measured $\fatq_{\rm SDW}$ range, suggesting that SDW order in NbFe$_2$ has a more subtle origin. The observed dependence of $q_{\rm SDW}$ (Fig.~\ref{figure_3}c), which decreases as the FM state is approached, indicates that $\chi_q$ is strongly modified by order parameter fluctuations on the threshold of FM.

\noindent The close connection between modulated and uniform magnetic order in
NbFe$_2$ is striking. It can be modelled effectively with the help
of a Landau expansion of the free energy in terms of the two order
parameters $M$ (for FM) and $P$ (for SDW order) \cite{fri18a}, which in zero
field reduces to
\begin{equation}
F/\mu_0 = \frac{a}{2}M^2 +\frac{b}{4}M^4
+\frac{\alpha}{2}P^2 + \frac{\beta}{4}P^4 + \frac{\eta}{2}P^2 M^2
\label{LandauTheory}
\end{equation}
The use of this model is corroborated in this study by the observation that $q_{\rm SDW}$ does not go to zero but stays finite at the SDW-FM transition. While the parameters $a$ and $b$ can be obtained from magnetization measurements, this is not possible for the parameters $\alpha$,
$\beta$ and $\eta$. They can, however, be determined from the SDW
ordered moment $P$ observed in our neutron scattering study.  We make
use of the Landau theory result that $P^2 = -\alpha/\beta$ and that
within the SDW phase the intercept $a^*$ and slope $b^*$ of the Arrott
plot $H/M$ vs. $M^2$, which can be determined in bulk magnetization
measurements, become $a^* = a-\alpha \eta/\beta$ and $b^* =
b-\eta^2/\beta$ \cite{fri18a}. Writing $a = a_1(T-T_0)$ and $\alpha = \alpha_1
(T-T_N)$, we find that the parameters characterising the SDW order
vary slowly with composition and are of similar magnitude as those
characterising the FM order: $\alpha_1/\alpha \simeq \beta / b \simeq
2\pm 1$ and that the coupling parameter $\eta \simeq \beta$ \cite{nik19a_sup}. This implies that the ratio $\eta/\sqrt{\beta b}>1$ throughout, which rules out phase coexistence within the two-parameter Landau theory, in agreement with experimental observations. Most importantly, the parameter values point towards a common microscopic origin of both types of magnetic order and puts strong constraints on a microscopic theory of magnetism in NbFe$_2$.

A number of theoretical studies \cite{bel97a,voj99a,chu04a,bra16a} have noted that the FM quantum critical point is unlikely to be observed in clean, 3D ferromagnets, and will instead either be avoided by a change of the magnetic transition from second order to first order or masked by the emergence of long-wavelength SDW. Both scenarios can be attributed to nonanalytic terms in the free energy associated with soft modes. These contribute a singular $q$-dependence of the form $q^2\ln q$ to $\chi_q^{-1}$ and thereby produce an intrinsic tendency towards long-wavelength modulated order as the FM QCP is approached. An alternative approach \cite{kar12a} arrived at similar conclusions by considering the contribution of order parameter fluctuations to the free energy, when different ordered states were imposed, causing differences in the phase space available to critical fluctuations of FM and modulated magnetic order. Helical order in the local moment system PrPtAl \cite{abd15a} has recently been presented as a likely manifestation of this scenario, but a demonstration in a band magnet is still outstanding. 

New neutron data demonstrate that SDW order emerges near the border of ferromagnetism in the C14 Laves phase system Nb$_{1-y}$Fe$_{2+y}$, burying an underlying FM QCP. This suggests that the SDW order in Nb$_{1-y}$Fe$_{2+y}$ is caused by an intrinsic instability of a ferromagnetic quantum critical point to modulated magnetic order, which has been postulated on the basis of fundamental considerations but has not before been detected in a band magnet.

%%%%%%%%%%%%%%%%%%%%%%%%%%%%%%%%%%%%%%%%%%%%%%%%%%%%%%%%%%%%%%%%%%%%%%%%%

This work is based upon experiments performed at the PANDA instrument operated by JCNS at the Heinz Maier-Leibnitz Zentrum (MLZ), Garching, Germany, and at the 4F2 instrument at LLB, CEA-Saclay, France. We acknowledge support by the EPSRC through grant EP/K012894/1 and by the German Science Foundation (DFG) through FOR 960 (CP) and SFB/TR 80 (CP). This research project has also been supported by the European Comission under the 7th Framework Programme through the 'Research Infrastructures' action of the 'Capacities' Programme, Contract No: CP-CSA\_INFRA-2008-1.1.1 Number 226507-NMI3. We thank S. Friedemann, M. Brando, and S. S\"ullow for helpful discussions.

\vspace{-0.2cm}

%\bibliography{refs}

%\vspace{-0.2cm}

\end{document}